\documentclass[prb,showpacs,twocolumn,superscriptaddress,showkeys,floatfix]{revtex4}

\usepackage{graphicx}
\usepackage{color}
\usepackage{dcolumn}
\usepackage{amsmath}
\usepackage{amssymb}
\usepackage[colorlinks=false]{hyperref}

\input{MyMc} 

\newcommand{\affTU}{\affiliation{%
  Physikalisches Institut -- Center for Collective Quantum Phenomena and their Applications in LISA+,
  Universit\"at T\"ubingen,
  Auf der Morgenstelle 14,
  D-72076 T\"ubingen, Germany
}}
\newcommand{\affKA}{\affiliation{%
  Institut f\"ur Mikro- und Nanoelektronische Systeme, 
  Karlsruher Institut f\"ur Technologie, 
  Hertzstra\ss e 16, D-76187 Karlsruhe, Germany
}}

\begin{document}

\title{Spectroscopy of a fractional Josephson vortex molecule}

\author{U. Kienzle}
\email{uta.kienzle@uni-tuebingen.de}
\affTU
\author{J. M. Meckbach} 
\affKA
\author{K. Buckenmaier} 
\author{T. Gaber}
\author{H. Sickinger} 
\affTU
\author{Ch. Kaiser} 
\author{K. Ilin} 
\author{M. Siegel} 
\affKA
\author{D. Koelle} 
\author{R. Kleiner} 
\author{E. Goldobin}
\affTU


\date{\today}

\begin{abstract}
In long Josephson junctions with multiple discontinuities of the Josephson phase, fractional vortex molecules are spontaneously formed. At each discontinuity point a fractional Josephson vortex carrying a magnetic flux $|\Phi|<\Phi_0$, $\Phi_0\approx 2.07\times 10^{-15}$ Wb being the magnetic flux quantum, is pinned. Each vortex has an oscillatory eigenmode with a frequency that depends on $\Phi/\Phi_0$ and lies inside the plasma gap.
We experimentally investigate the dependence of the eigenfrequencies of a two-vortex molecule on the distance between the vortices, on their topological charge $\wp=2\pi\Phi/\Phi_0$ and on the bias current $\gamma$ applied to the Josephson junction. We find that with decreasing distance between vortices, a splitting of the eigenfrequencies occurs, that corresponds to the emergence of collective oscillatory modes of both vortices.  We use a resonant microwave spectroscopy technique and find good agreement between experimental results and theoretical predictions.

\end{abstract}

\maketitle
\section{Introduction}

During the last decade the physics of fractional Josephson vortices has attracted a lot of attention.\cite{Kirtley:SF:HTSGB, Kirtley:SF:T-dep, Smilde:ZigzagPRL, Hilgenkamp:zigzag:SF, Goldobin:Art-0-pi, Weides06a, Buckenmaier:2007:ExpEigenFreq, Dewes08} In contrast to the well known fluxons,\cite{Likharev86} fractional vortices carry the magnetic flux of $|\Phi|<\Phi_0$. 
Here we focus on the fractional vortices that emerge at the discontinuities of the Josephson phase\cite{Bulaevskii:0-pi-LJJ, Xu:SF-shape, Goldobin:SF-Shape} in long Josephson junctions (LJJ) and disregard all other types, \eg, splintered vortices.\cite{Mints02,Buzdin:2003:phi-LJJ,Goldobin07} The phase discontinuity $\kappa$ can be created by two tiny (compared to the Josephson penetration depth $\lambda_J$) current injectors,\cite{Ustinov:2002:ALJJ:InsFluxon, Goldobin:SF-Shape, Goldobin:Art-0-pi} so that $\kappa\propto I_\mathrm{inj}$, where $I_\mathrm{inj}$ is the current sent through the injectors.

A well known property of a long Josephson junction is the presence of a gap in the spectrum of electromagnetic waves (plasma waves) --- the so-called plasma gap, which ranges from 0 to the plasma frequency $\omega_p$. The value of $\omega_p \propto \sqrt{j_c}$ is $\sim 10\ldots1000\units{GHz}$, where $j_c$ is the  critical current density of the junction. In a LJJ with a fractional vortex pinned at the phase discontinuity, the vortex has an eigenfrequency $\omega_0$ which is situated within the plasma gap (localized mode with a wave number $k=0$). The value of $\omega_0(\wp, \gamma)$ depends on the topological charge $\wp$ of the vortex and the normalized bias current $\gamma$ applied to the LJJ.\cite{Goldobin:2KappaEigenModes}

The topological charge of a single fractional vortex in an infinite LJJ with a $\kappa$ discontinuity of the phase is defined as $\wp=\phi(+\infty)-\phi(-\infty)-\kappa$, where $\phi(x)$ is the Josephson phase, that is $\kappa$-discontinuous at one point. In simple words, $\wp$ is a continuous advance of the phase when one goes from $x=-\infty$ to $x=+\infty$. Among all possible combinations of the phases at $x=\pm\infty$ only two irreducible combinations are stable\cite{Goldobin:2KappaGroundStates}: the direct vortex with $\wp = -\kappa$ and a complementary vortex with $\wp =-\kappa+2\pi\,\mathrm{sgn}(\kappa)$. We assume that $\left|\kappa\right| \leq2\pi$, otherwise the structure either cannot be constructed (not a solution of the sine-Gordon equation) or is unstable --- it splits into one or several fluxons and a fractional vortex with $|\wp|\leq2\pi$.\cite{Goldobin:2KappaGroundStates} For $\wp\rightarrow 0$ there is a smooth transition to the flat phase state $\phi(x)=0$ of the LJJ, while for $|\wp|\rightarrow2\pi$ a(n) (anti)fluxon is formed. In contrast to the fractional vortex this fluxon is not pinned anymore and can move freely.\cite{Likharev86}

The eigenfrequency of a single fractional vortex as a function of $\wp$ and $\gamma$ was already measured experimentally\cite{Buckenmaier:2007:ExpEigenFreq,Gaber07} using microwave spectroscopy. Good agreement with numerical calculations was found.

Nowadays more complex fractional vortex systems, \eg fractional vortex molecules consisting of two, three, \etc vortices can be created and studied experimentally.\cite{Dewes08} In particular, two-vortex molecules with degenerate ground states\cite{Goldobin:2KappaGroundStates} are proposed as a candidate for a macroscopic qubit.\cite{Goldobin:2005:MQC-2SFs} In such a molecule, the coupling between vortices results in shifting/splitting of their eigenfrequency,\cite{Goldobin:2KappaEigenModes} very similar to the splitting of electronic energy levels in a real molecule when it is formed from atoms.

In this paper we experimentally investigate the eigenfrequencies in two types of symmetric two-vortex molecules: one with parallel (ferromagnetic) configuration and the topological charges $(\wp,\wp)$ and one with antiparallel (antiferromagnetic) order and the topological charges $(\wp,-\wp)$. We use a spectroscopic measurement technique\cite{Buckenmaier:2007:ExpEigenFreq} to determine the eigenmodes of the molecule as a function of applied bias current and topological charges. 

The paper is organized as follows. In Sec.~\ref{Sec:Theory} we present the underlaying theory and present numerically calculated ground states of molecules as well as their eigenfrequencies. Experimental results are presented and discussed in Sec.~\ref{Sec:Samples}. Section~\ref{Sec:Conclusions} concludes the work.

\section{Theory}
\label{Sec:Theory}

The perturbed sine-Gordon equation, which describes the dynamics of the Josephson phase  is given by\cite{Goldobin:SF-Shape}
\begin{equation}
  \phi_{xx}-\phi_{tt}-\sin(\phi) = \alpha\phi_t-\gamma(x) + \theta_{xx}(x),
  \label{sineGordon}
\end{equation}
where subscripts $x$ and $t$ denote derivatives with respect to space and time and the step function $\theta(x)$ describes the position of discontinuities. In the case of one $\kappa$ discontinuity located at $x=0$, $\theta(x)=\kappa\Heavyside(x)$, where $\Heavyside(x)$ is a Heaviside step function. In Eq.~\eqref{sineGordon} the coordinate $x$ and time $t$ are normalized to the Josephson penetration depth $\lambda_J$ and the inverse plasma frequency $\omega^{-1}_{p}$, respectively. $\alpha=1/\sqrt{\beta_c}$ is the dimensionless damping parameter, $\beta_c$ is the Stewart-McCumber parameter. The bias current density $\gamma$ is normalized to the critical current density $j_c$ of the junction. To solve Eq.~\eqref{sineGordon} numerically we introduce a new continuous phase\cite{Goldobin:SF-Shape} $\mu(x,t)=\phi(x,t)-\theta(x)$ and solve the resulting \emph{smooth} equation for $\mu(x,t)$.

The eigenfrequency of a single vortex with a topological charge $\wp$ in an unbiased ($\gamma=0$) infinite LJJ without damping ($\alpha=0$) is given by \cite{Goldobin:2KappaEigenModes}
\begin{equation}
  \omega_0(\wp)= \omega_p
  \sqrt{\frac12\cos\frac{\wp}{4}\left(\cos\frac{\wp}{4}+\sqrt{4-3\cos^2\frac{\wp}{4}}\right)}
  .
  \label{Eq:SG}
\end{equation}
In case of $\gamma\neq0$ no analytical expression has been found and therefore the eigenfrequency has to be calculated numerically. A good approximation to the numerical results is given by
\begin{equation}
  \omega_0(\wp,\gamma)\simeq\omega_0(\wp,0)\cdot\sqrt[4]{1-\left(\frac{\gamma}{\gamma_c}\right)^2},
  \label{Eq:omegagamma}
\end{equation}
where
\begin{equation}
  \gamma_c(\wp)= \left|\frac{\sin(\wp/2)}{\wp/2}\right|,\;|\wp|\leq2\pi
  \label{Eq:sincfkt}
\end{equation}
is the  depinning current\cite{Malomed:2004:ALJJ:Ic(Iinj),Goldobin:F-SF} of the vortex. For bias currents exceeding $\gamma_c$, the fractional vortex flips into its complementary partner, thereby emitting a fluxon.
 

If one considers two discontinuities with the same absolute value of $|\kappa|$ that are situated at $x_{1,2}=\pm a/2$ ($a$ is the distance between the discontinuities) in an infinite LJJ, four irreducible vortex configurations are possible.\cite{Goldobin:2KappaGroundStates} In the case of equal discontinuities, \ie $(-\kappa , -\kappa)$, there exists the symmetric parallel (ferromagnetically) ordered (FM) vortex molecule with  $(\wp_1,\wp_2)=(+\kappa, +\kappa)$ and the asymmetric antiparallel (antiferromagnetically) ordered (AFM) vortex molecule with $(\wp_1,\wp_2)=(+\kappa ,+\kappa-2\pi)$. For discontinuities of different signs, \ie $(+\kappa , -\kappa)$, one gets either the symmetric antiparallel molecule with  $(\wp_1,\wp_2)=(-\kappa,+\kappa)$ or the asymmetric parallel vortex molecule with $(\wp_1,\wp_2)=(-\kappa+2\pi,+\kappa)$.

In the experiments described below we use annular LJJs\cite{Davidson85} with two discontinuities at a distance $a=l/2$ from each other, where $l=2\pi R/ \lambda_J$ is the  mean circumference of the junction normalized to $\lambda_J$. There are certain advantages of this particular geometry. First, due to a particular shape of the electrodes, annular junctions have intrinsically a much more homogeneous bias current density than linear LJJs,\cite{Martuciello98} which is important to measure $\omega_0(\gamma)$. Second, any real LJJ has a finite length, which results in a somewhat modified phase profile and eigenfrequency spectrum of the vortex molecules. The boundary conditions of the junction thereby impose an additional level of symmetry/ordering on the molecule. In an annulus this additional symmetry is always the same as the internal configuration of the molecule, FM or AFM.\footnote{The phase profile of a two-vortex molecule in a finite LJJ can be constructed by employing vortex images using the boundaries of the junction as mirrors.\cite{Goldobin:Art-0-pi} In linear geometry the phase resembles a piece of length $l$ of an infinite chain of AFM ordered vortex pairs, even for a FM ordered molecule. Annular geometry, however, results in a FM order of vortex pairs, thereby conserving the molecules inner configuration.} 

In an annular junction each vortex has the same neighbor on both sides (right and left) so that magnetic field profiles and eigenfrequencies are somewhat different from the results obtained earlier for an infinite LJJ.\cite{Goldobin:2KappaGroundStates} This is especially true for short JJs. In principle, there are many eigenmodes $\omega_0<\omega_1<\ldots$ enumerated according to their frequency. In the following we use $\omega_+$ to denote the eigenfrequency of the vortex in-phase oscillations, while $\omega_-$ denotes the oscillation mode with a phase shift of $\pi$. They correspond to the two lowest eigenmodes $\omega_{0}$ and $\omega_1$, \ie $\omega_\pm=\omega_0$ or $\omega_1$. Depending on the configuration of the molecule $\omega_+$ can be larger or smaller than $\omega_-$. In general, the eigenfrequencies of a molecule can only be calculated numerically.\cite{Goldstkjj06} Due to the vast number of possible molecule configurations we, in the following, limit ourselves to the \emph{symmetric} configurations and discuss them in more detail. 

\subsection{Parallel (FM) molecule}
\begin{figure}[ht] 
  \begin{center}
    \includegraphics{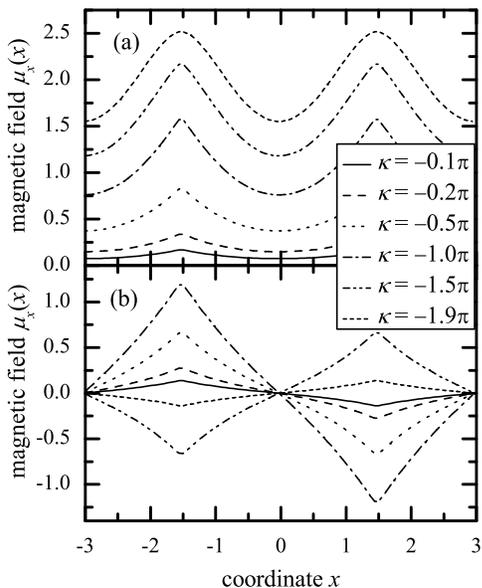}
  \end{center}
  \caption{%
    Magnetic field profile $\mu_x(x)$ in an annular LJJ of length $l=6$ and $a=3$ for different values of $\kappa$: (a) parallel configuration, (b) antiparallel configuration.
  }
  \label{Fig:MagnField}
\end{figure}

The magnetic field $\propto\mu_x(x)$ of a molecule consisting of two parallel ordered vortices pinned at $(-\kappa,-\kappa)$ discontinuities is exemplarily shown in Fig.~\ref{Fig:MagnField}(a) for an annular LJJ of length $l=6$. The vortex distance is half of the junction length. No bias current is applied.
Then, the FM configuration is stable for $|\kappa|<2\pi$ and for all lengths due to annular boundary conditions and $a=l/2$, \emph{cf.} Refs.~\onlinecite{Goldobin:2KappaEigenModes} and \onlinecite{Goldobin:2KappaGroundStates} for more general cases. For $|\kappa|\geq2\pi$, a single direct vortex with $|\wp|\ge2\pi$ is not stable anymore. Instead, at each discontinuity a(n) (anti)fluxon and a $(\kappa-2\pi\sgn(\kappa))$-vortex (all of the same polarity)\cite{Goldobin:2KappaGroundStates} is formed. 

\begin{figure} 
  \begin{center}
    \includegraphics{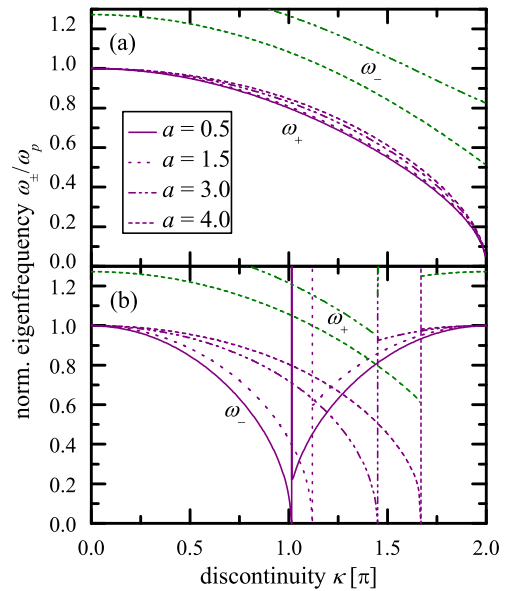}
  \end{center}
  \caption{%
    (Color online) Numerically calculated lowest eigenmode(s) of a fractional vortex molecule in an annular LJJ for different distances $a$ between the vortices: (a) parallel configuration, (b) antiparallel configuration. The junction length $l$ is twice the distance $a$.}
  \label{Fig:EigenFreq}
\end{figure}

The eigenfrequencies $\omega_{\pm}$ of a FM ordered molecule are shown in Fig.~\ref{Fig:EigenFreq}(a) as a function of $\kappa$ for different vortex distances. 
Here, $\omega_+<\omega_-$. The dependence of $\omega_{\pm}(a)$ is shown in Fig.~\ref{Fig:EigenFreqOfa} for two molecules with specific values of the discontinuities $\kappa_i=\kappa$. For comparison, the eigenfrequency of a single vortex in an \emph{infinitely} long JJ is also shown. 
As can be seen the frequency splitting increases for smaller $a$ (larger coupling). However, the divergence of the out-of-phase mode $\omega_-=\omega_1$ for $a\to0$ is exclusively related to the fact, that also $l=2a\to0$ and therefore $\omega_1 \propto 1/l\to\infty$ for $l\ll1$.

\subsection{Antiparallel (AFM) molecule}
\label{SSec:AFM}
Figure \ref{Fig:MagnField}(b) shows the magnetic field $\propto\mu_x(x)$ for the antiparallel molecule pinned at $(-\kappa, +\kappa)$ discontinuities in an annular LJJ of $l=6$.
In contrast to the parallel configuration there exists a critical value $\kappa_c$ ($\pi<\kappa_c<2\pi$), above which the vortices are not stable anymore: after exchanging one $\Phi_0$ the $(+\kappa,-\kappa)$-molecule turns into a complementary one $(+\kappa-2\pi,-\kappa+2\pi)$. The value of $\kappa_c$ depends on $a$ and can in general only be calculated numerically. For an AFM molecule in an infinitely long JJ $\kappa_c$ was calculated in Ref.~\onlinecite{Goldobin:2KappaGroundStates}. Note that for $(2\pi-\kappa_c)< |\kappa|<\kappa_c$ both states $(+\kappa,-\kappa)$ and $(+\kappa-2\pi,-\kappa+2\pi)$ are stable and that for $\kappa\neq\pi$ these solutions have different energies.

The eigenfrequencies $\omega_{\pm}(\kappa)$ of the two lowest modes of the antiparallel configuration are shown in Fig.~\ref{Fig:EigenFreq}(b). The vertical lines describe the critical discontinuity $\kappa_c$ for different vortex distances $a$. In contrast to the parallel configuration, here $\omega_+>\omega_-$. 

\begin{figure}[htb] 
  \begin{center}
    \includegraphics{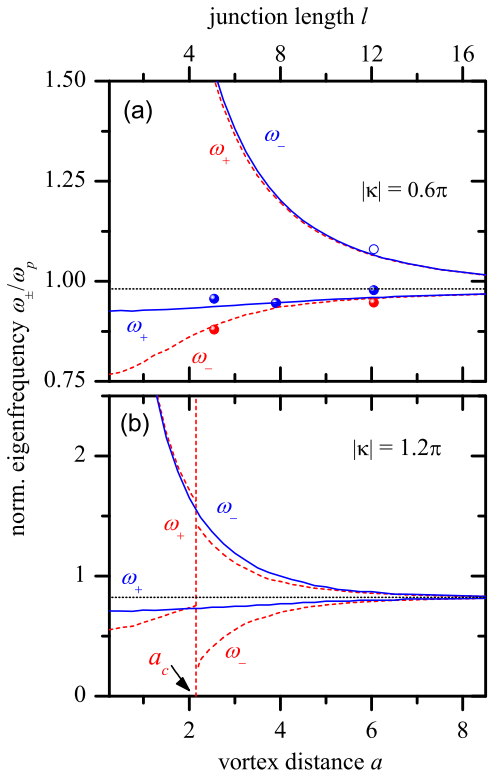}
  \end{center}
  \caption{%
    (Color online) The two lowest eigenfrequencies $\omega_\pm(\kappa)$ of a symmetric vortex molecule in an annular LJJ with length $l=2a$ for different distances $a$ between the vortices: (a) $|\kappa|=0.6\pi$ and (b) $|\kappa|=1.2\pi$. Solid (dashed) lines correspond to the FM (AFM) configuration. For comparison the eigenfrequency of a single vortex in an \emph{infinitely} long JJ is also shown (dotted lines). Full (open) symbols correspond to fitted values of $\omega_0$ ($\omega_1$) from experimental data, see text.
  }
  \label{Fig:EigenFreqOfa}
\end{figure}

The dependence of the eigenfrequencies of two antiparallel vortices for fixed $\kappa$ values as a function of the distance $a$ is shown in Fig.~\ref{Fig:EigenFreqOfa}. As with the parallel configuration the frequency splitting increases when $a$ decreases. One has to recognize, that for $\kappa=1.2\pi$ the critical vortex distance is $a_c\approx2.15$. Therefore, the vortex state $(+\kappa,-\kappa)$ becomes unstable at $a \to a_c$ and the frequency $\omega_- =\omega_0\to 0$. For $a<a_c$, $\omega_\pm$ correspond to the complementary state $(\kappa-2\pi,2\pi-\kappa)$, which is stable for all $a$. In comparison to the lowest eigenfrequency of the parallel configuration, this mode has a smaller frequency and this difference increases as $a \to 0$.

For a biased molecule ($\gamma\neq0$) the approximation \eqref{Eq:omegagamma} can also be used to estimate $\omega_\pm(\wp,\gamma)$.

\section{Experiments}
\label{Sec:Samples}
\begin{figure}
  \centering \includegraphics{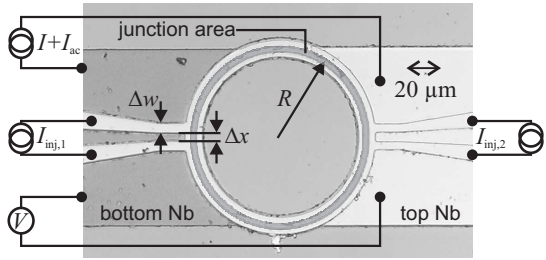}
  \caption{%
    Optical image (top view) of sample 2: annular LJJ with two pairs of injectors. The dc bias $I$ is superimposed with a microwave induced ac current $I_\mathrm{ac}$, that is capacitively coupled into the bias leads. $I_{\mathrm{inj},1}$ is applied to the left (top) pair of injectors, $I_{\mathrm{inj},2}$ to the right (bottom) pair respectively.
   }
   \label{Fig:sample}
\end{figure}

For experiments we used underdamped long Nb-AlO$_x$-Nb annular tunnel junctions fabricated using standard trilayer technology.\cite{Kaiser11} Each JJ was equipped with two pairs of current injectors as shown in Fig.~\ref{Fig:sample}.
Note that injector pairs are placed opposite to each other and directly on top/below the bias electrodes of the junction, as earlier measurements showed this layout to give best agreement of the critical current dependence $I_c(I_\mathrm{inj})$ with theory, \ie Eq.~(\ref{Eq:sincfkt}). As the JJs have been fabricated in an overlap geometry, their static and dynamic properties are modified by the overlap region of top and bottom electrode, the so-called idle region.\cite{Caputo:1994:IdleReg:DressFluxon,Caputo:1996:IdleRgn:Dynamics,Caputo:1999:IdleReg:EffSine-Gordon,Lee:1991:IdleReg:LinRes,Lee:1992:IdleReg:Dispersion,Flytzanis:2000:IdleReg:FluxonDyn} Idle region effects can be taken into account by a renormalization of the junction length, \ie by introducing an effective Josephson penetration depth $\lambda_{J,\mathrm{eff}}>\lambda_J$. However, we find that the usual correction\cite{Caputo:1999:IdleReg:EffSine-Gordon} overestimates the idle region effect. Table~\ref{Tab:samples} summarizes the properties of the measured junctions, listing Josephson penetration depths and normalized junction lengths, with and without idle region corrections, as well as the normalized lengths we found to give best agreement with theory. 
\begin{table}
\centering
\begin{tabular}{|l|l|l|l|l|l|l|l|l|}
\hline
\multicolumn{1}{|c|}{\#} & \multicolumn{1}{c|}{$R$} & \multicolumn{1}{c|}{$j_c$} & \multicolumn{1}{c|}{$\lambda_J$} &      \multicolumn{1}{c|}{$l_b$} & \multicolumn{1}{c|}{$\lambda_{J,\mathrm{eff}}$} & \multicolumn{1}{c|}{$l_{\mathrm{eff}}$} & \multicolumn{1}{c|}{$l$} & \multicolumn{1}{c|}{$\omega_p/2\pi$} \\ 
\multicolumn{1}{|c|}{} & \multicolumn{1}{c|}{[$\mu$m]} & \multicolumn{1}{c|}{[A/cm$^2$]} & \multicolumn{1}{c|}{[$\mu$m]} & \multicolumn{1}{c|}{} & \multicolumn{1}{c|}{[$\mu$m]} & \multicolumn{1}{c|}{} & \multicolumn{1}{c|}{} & \multicolumn{1}{c|}{[GHz]} \\ 
\hline
\multicolumn{1}{|c|}{1} & \multicolumn{1}{c|}{30} & \multicolumn{1}{c|}{112} & \multicolumn{1}{c|}{35.7} & \multicolumn{1}{c|}{5.3} & \multicolumn{1}{c|}{41.4} & \multicolumn{1}{c|}{4.6} & \multicolumn{1}{c|}{5.1} & \multicolumn{1}{c|}{45.1} \\ 
\hline
\multicolumn{1}{|c|}{2} & \multicolumn{1}{c|}{50} & \multicolumn{1}{c|}{90} & \multicolumn{1}{c|}{39.8} & \multicolumn{1}{c|}{7.9} & \multicolumn{1}{c|}{46.1} & \multicolumn{1}{c|}{6.8} & \multicolumn{1}{c|}{7.8} & \multicolumn{1}{c|}{44} \\ 
\hline
\multicolumn{1}{|c|}{3} & \multicolumn{1}{c|}{80} & \multicolumn{1}{c|}{77} & \multicolumn{1}{c|}{40.9} & \multicolumn{1}{c|}{12.3} & \multicolumn{1}{c|}{43.7} & \multicolumn{1}{c|}{11.4} & \multicolumn{1}{c|}{12.1} & \multicolumn{1}{c|}{36} \\ 
\hline
\end{tabular}
\caption{Sample properties measured at $T=4.2\,\mathrm{K}$. Junction widths $w$ and injector distances $\Delta x$ are $5\,\mathrm{\mu m}$, injector widths $\Delta w$ are $5\,\mathrm{\mu m}$, except for sample 3, where $\Delta w=2\,\mathrm{\mu m}$. $l_\mathrm{eff}=2\pi R/\lambda_{J,\mathrm{eff}}$ and $l_b=2\pi R/\lambda_J$ are the normalized junction length with and without idle region corrections and $R$ is the radius of the junction. $l$ is the length which gives best agreement between measurement and theory. $\omega_p$ is the plasma frequency obtained from spectroscopic measurements.}
\label{Tab:samples}
\end{table}
%

The measurements shown in this paper have been performed at $T=4.2\units{K}$ with the resonant vortex excitation technique already used to determine the eigenfrequency of a single fractional vortex.\cite{Buckenmaier:2007:ExpEigenFreq} In brief, we continuously apply microwaves with fixed frequency $\omega_\mathrm{ext}<\omega_0(\wp)$ and fixed power to the junction while ramping up the bias current $\gamma$. As $\omega_i(\wp,\gamma)$ decreases with increasing $\gamma$, \cf Eq.~\eqref{Eq:omegagamma}, the resonance condition  $\omega_0(\wp,\gamma_r)\simeq \omega_\mathrm{ext}$ is achieved for some $\gamma_r$. Then, depending on applied microwave power and the internal damping of the junction, the vortex molecule may flip and switch the junction into its resistive state.\cite{Fistul00} If, however, the junction remains in the zero-voltage state it switches to the non-zero voltage state by means of thermal escape when the bias approaches the fluctuation-free critical current $\gamma_c$. By repeatedly measuring the switching current and systematically scanning microwave power and frequency we can map out the resonances corresponding to the eigenmodes of the molecule. As the JJ itself is a non-linear oscillator, resonances shift towards lower frequencies with increasing microwave power.\cite{Likharev86, Fistul00, Gronbech04a, Gronbech04b, Blackburn10} This is especially relevant at small $\gamma$ where the energy barrier between the zero and the non-zero voltage state is still large and high microwave power is necessary for switching. We therefore determine $\gamma_r$ from the measurements with the lowest power to keep non-linear effects to a minimum.

\begin{figure}[htb] 
  \centering \includegraphics{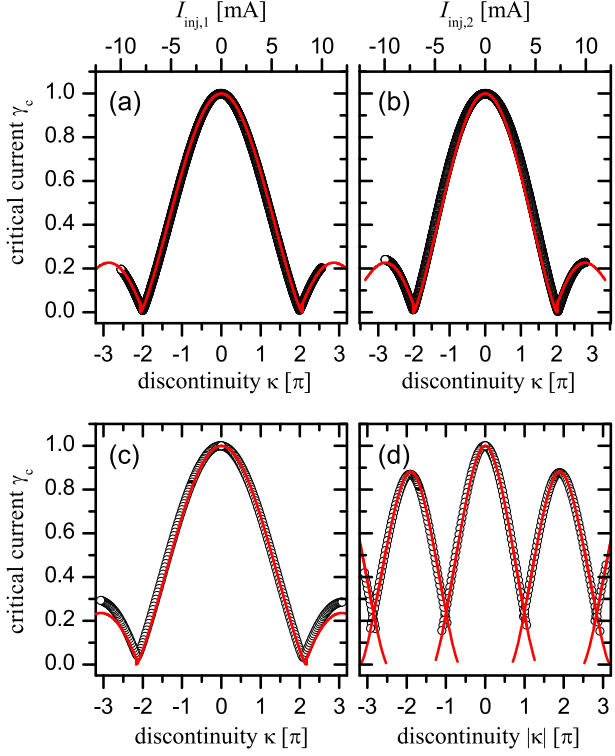}
  \caption{%
    (Color online) Measured (symbols) and numerically calculated (red line) critical current $\gamma_c=I_c(I_{\mathrm{inj},i})/I_c(0)$ as a function of injector current. Single vortex cases with either (a) injector current $I_{\mathrm{inj},1}$ or (b) $I_{\mathrm{inj},2}$ applied. (c) FM and (d) AFM molecule cases corresponding to  $(+\kappa,+\kappa)$ and $(+\kappa,-\kappa)$ discontinuities. Data is shown for sample 1. For the vortex molecule injector currents $I_{\mathrm{inj},i}=\pm\kappa|I_{\mathrm{inj},i}^\mathrm{min}|/2\pi$ were applied simultaneously.
   }
   \label{Fig:IcInj}
\end{figure}

%
%
Before spectroscopy was performed, injectors were calibrated. For this we first measured the critical current $I_c$ of the junction as a function of the current through each injector pair $I_{\mathrm{inj},1}$ and $I_{\mathrm{inj},2}$ individually. From Eq.~(\ref{Eq:sincfkt}) and $\kappa\propto I_\mathrm{inj}$ we then determined the injector currents $I_{\mathrm{inj},i}^\mathrm{min}$ necessary to create a discontinuity of $\kappa_i = 2\pi$, where $I_c(I_{\mathrm{inj},i})$ have the cusp like first minima. Figure~\ref{Fig:IcInj}(a) and (b) show the measured $I_c(I_{\mathrm{inj},i})$ dependences of sample 1, as well as the fitted theoretical curves. 
Although both injector pairs have nominally the same width and distance, the $I_{\mathrm{inj},i}^\mathrm{min}$ are usually somewhat different. In sample 3 the inequality was introduced intentionally, as we deliberately cut the top wiring layer between the injectors to increase the coupling of $I_{\mathrm{inj},1}$ to the junction. However, in samples 1 and 2, both injector pairs are identical by design. Here, we believe, the deviations are due to lithographical inaccuracy and different thicknesses of the top and bottom Nb layer. 
To create \emph{symmetric} vortex molecules we henceforth applied the injector currents $I_{\mathrm{inj},i}=\pm\kappa |I_{\mathrm{inj},i}^\mathrm{min}|/2\pi$ simultaneously. 

Figure~\ref{Fig:IcInj}(c) shows the measured and the numerically calculated $\gamma_c=I_c(\kappa)/I_c(0)$ dependence of a symmetric FM vortex molecule pinned at $(\kappa,\kappa)$ discontinuities. As can be seen there is a good agreement between experiment and theory. Note that $\gamma_c(+\kappa,+\kappa)$ looks qualitatively similar to $\gamma_c(\kappa)$ for a single vortex,\cite{Malomed:2004:ALJJ:Ic(Iinj)} yet the first minimum positions are slightly different: $\gamma_c^\mathrm{min}$ of a symmetric FM vortex molecule in an annular JJ with a \emph{finite} width $\Delta x$, $\Delta w$ of injector pairs corresponds to a $\kappa_i$ value, which is slightly larger than $2\pi$.

The $\gamma_c(+\kappa,-\kappa)$ curve for the AFM molecule is depicted in Fig.~\ref{Fig:IcInj}(d). Again, experimental data and numerical calculations are in good agreement. In contrast to the FM configuration, here the $\gamma_c$-minima correspond to $|\kappa|\simeq(2n+1)\pi$ instead of $\kappa\simeq 2n\pi$, \cf Eq.~(\ref{Eq:sincfkt}). Note that the $\gamma_c(+\kappa,-\kappa)$ curve is identical to the $\gamma_c(\kappa)$ dependence of a \emph{linear} LJJ of length $l/2$ with a \emph{single} central vortex.\cite{Gaber05} As already mentioned in Sec.~\ref{SSec:AFM}, the AFM molecule exhibits regions of bistability of the direct and the complementary vortex states around $|\kappa|=(2n+1)\pi$. Between $2\pi-\kappa_c(a)<\kappa<\kappa_c(a)$ the critical current of both vortex states can be traced out,\cite{Dewes08} see Fig.~\ref{Fig:IcInj}(d). In AFM configuration the $\gamma_c(+\kappa,-\kappa)$ pattern sensitively depends on junction length as well as on injector size\cite{Gaber05}: For infinitesimally small injectors, the $\gamma_c(+\kappa,-\kappa)$ is $2\pi$ periodic with maxima at $\kappa=2\pi n$ and minima at $\kappa=(2n+1)\pi$. For finite size injectors, however, this is not the case: $\gamma_c(\pm 2\pi n,\mp 2\pi n)<\gamma_c(0,0)$ and the minimum positions shift to $\kappa>(2n+1)\pi$. In the FM case basically only $I_{\mathrm{inj},i}^\mathrm{min}$ depends on injector size --- the shape of the curve does not change. By carefully comparing the $I_c(I_{\mathrm{inj},i})/I_c(0)$ curves of the AFM configuration with numerical simulations we obtained the normalized junction length $l$, see Tab.~\ref{Tab:samples}, which we used for the subsequent calculation of eigenfrequencies.
\begin{figure} 
  \begin{center}
    \includegraphics{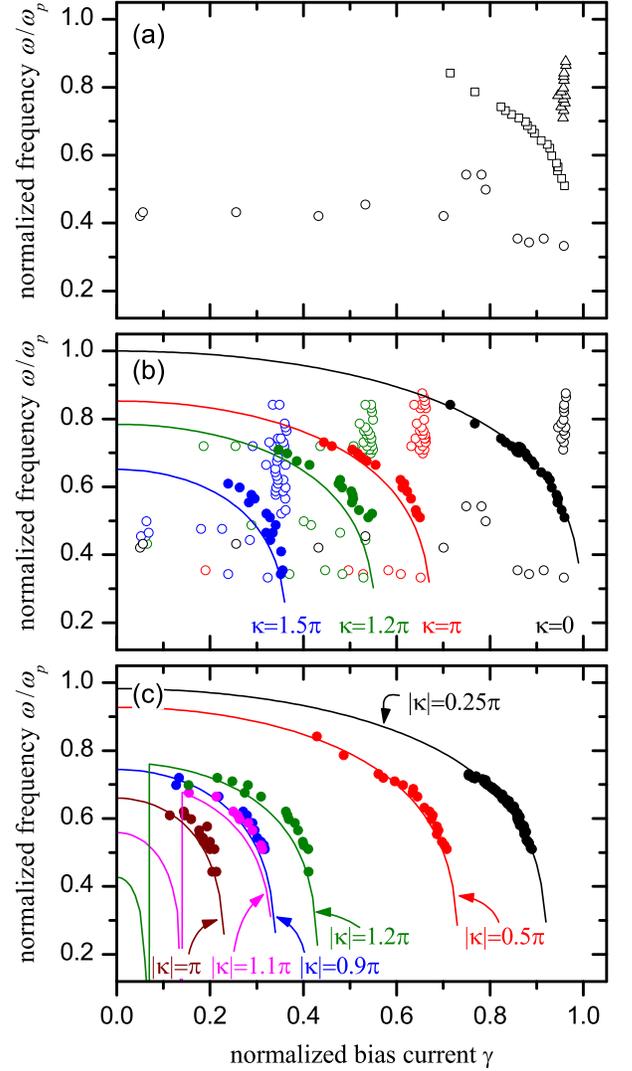}
  \end{center}
  \caption{%
    (Color online) Comparison of measured resonance peak positions (symbols) with the numerically calculated $\omega_0(\kappa,\gamma)$ dependences (solid lines). Data is shown for (a) the plasma frequency scan ($\kappa=0$), (b) the FM $(-\kappa,-\kappa)$ and (c) the AFM $(-\kappa,+\kappa)$ molecule configurations in sample 1. Symbols of same color correspond to the same $\kappa$ value. Symbols are explained in the text.
  }
  \label{Fig:FMandAFMS1}
\end{figure}

After calibration of injectors we spectroscopically mapped out the well-known dependence of the plasma frequency $\omega_0(0,\gamma)=\omega_p\sqrt[4]{1-\gamma^2}$ on applied bias current. 
Figure~\ref{Fig:FMandAFMS1}(a) shows exemplarily the results of the plasma frequency scan of sample 1. All measured resonance peak positions are plotted. Beside the desired plasma resonance (squares), additional resonances were observed. They do not correspond to a sub- or a superharmonic excitation of the plasma resonance, but can be subsumized into two groups. One set of resonances (triangles) occurs at high frequencies and close to the critical current $\gamma_c$ of the junction. These resonances are most likely the result of a microwave enhanced thermal escape. The second group of resonances (circles) seem to be erratically distributed over the full bias current range. Yet, a comparison with other spectroscopic measurements, \eg see Fig.~\ref{Fig:FMandAFMS1}(b), shows that they appear in rather small frequency bands, where the coupling of the microwave current to the junction is exceptionally high. We therefore believe that they are the result of a parasitic coupling of the junction to cavity modes of the sample box and the bias circuitry. Note, that in contrast to a typical spectroscopic measurement of the plasma frequency or the eigenfrequency of a single fractional vortex, \eg see Ref.~\onlinecite{Buckenmaier:2007:ExpEigenFreq}, we did not simply trace out the desired resonance curve by optimizing the measured frequency range and the applied microwave power. Instead, we systematically scanned over a large microwave power and frequency range to ensure, that no resonance is missed. Evidently, this measurement technique also reveals parasitic resonances, which are not of interest and must be removed in a manual post-selection process.
From the measured plasma resonance (squares) we extracted $\omega_p$ and the fluctuation-free critical current $I_0=j_c 2\pi R w$ of the junction, which then served as a reference for the following measurements on vortex molecules. $R$ is the junction radius, $w$ the junction width.
We also determined $\omega_p$ from measurements of the zero field steps\cite{Fulton73} on the current-voltage characteristics of the junction. After corrections due to idle region effects\cite{Lee:1991:IdleReg:LinRes,Lee:1992:IdleReg:Dispersion,Caputo:1996:IdleRgn:Dynamics} we typically find both values in agreement by better than 90\%.
%
%

Figure~\ref{Fig:FMandAFMS1}(b) shows all measured resonance peak positions of a set of FM ordered vortex molecules in sample 1. 
Also plotted are the corresponding numerically calculated $\omega_0(\kappa,\gamma)$ dependences. As sample 1 is rather short ($l=5.1$), $\omega_1=\omega_-$ is much larger than the highest frequency of our rf-generator and could therefore not be measured. As already discussed for the plasma frequency measurement, we partitioned the data --- eigenmode resonances are plotted with full symbols, whereas open symbols correspond to parasitic resonances. Beside a certain ambiguity of the data selection process, a good agreement with theory can clearly be seen. 
%
%
\begin{figure} 
  \begin{center}
    \includegraphics{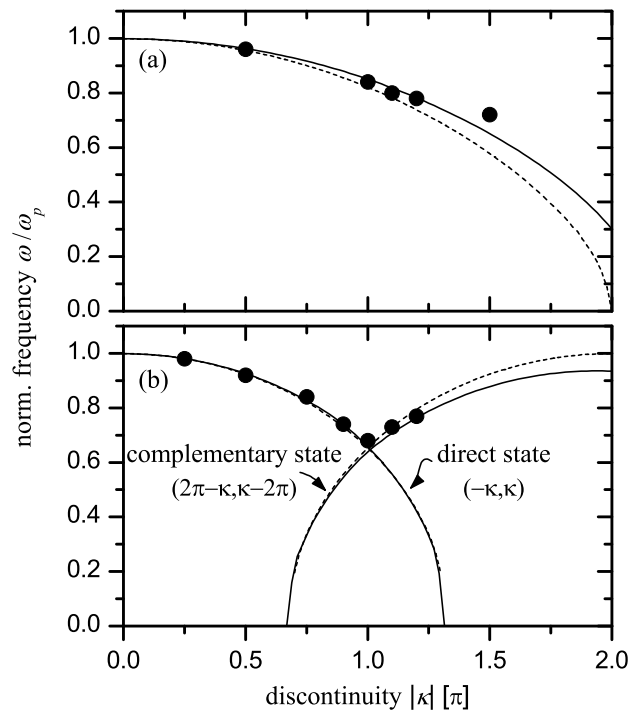}
  \end{center}
  \caption{%
  Comparison of extrapolated zero bias eigenfrequency (symbols) with the numerically calculated $\omega_0(\kappa,\gamma=0)$ dependence (lines) of a vortex molecule in sample 1. (a) FM order, (b) AFM order. Dashed lines correspond to ideal discontinuities, whereas solid lines take the finite injector size into account.
  }
  \label{Fig:EigenfreqG0S1}
\end{figure}
%
%

The measurement results of the more interesting AFM case are shown in Fig.~\ref{Fig:FMandAFMS1}(c). Here, injector currents were reset after each switching current measurement to ensure a well defined initial molecule state. Also, only eigenmode resonances are plotted. Again, there is a good agreement between experiment and theory. It is interesting to note, that for $\kappa>\pi$ almost no signature of the direct vortex state $(-\kappa,\kappa)$ can be seen. Instead, the measured resonances correspond to the complementary state $(2\pi-\kappa,\kappa-2\pi)$, which has a higher eigenfrequency $\omega_0$. As mentioned in Sec.~\ref{SSec:AFM}, for $\pi<\kappa<\kappa_c$ both the direct and the complementary molecule state are stable, yet have different energies and different critical currents, \cf Fig.~\ref{Fig:IcInj}(d). In case of sample 1, $\kappa_c\simeq 1.3\pi$. 
When a bias current is applied that exceeds $\gamma_c$ of the direct state, the molecule flips into the complementary state by exchange of one fluxon between the vortices. In contrast to the FM case however, here this transition is not automatically followed by the switching of the junction into a non-zero voltage state. So, as the experimental data indicates, already for $\kappa=1.1\pi$ the vortex flipping usually remains undetected and only the more stable complementary state is measured.
%
%
%
\begin{figure}[tb] 
  \begin{center}
    \includegraphics{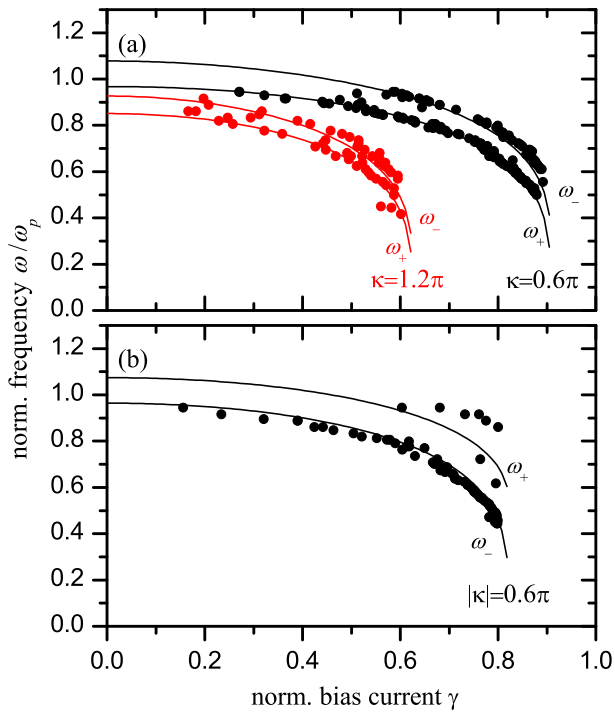}
  \end{center}
  \caption{%
    (Color online) Comparison of measured resonance peak positions (symbols) with the numerically calculated $\omega_\pm(\kappa,\gamma)$ dependences (solid lines). Data is shown for (a) FM and (b) AFM molecule configurations in sample 3. Same colors correspond to the same $\kappa$ value.
  }
  \label{Fig:FMandAFMS3}
\end{figure}

Using Eq.~(\ref{Eq:omegagamma}) we fitted the measured resonance peaks (full symbols) in Fig.~\ref{Fig:FMandAFMS1} leaving $\gamma_c(\kappa)$ and $\omega_0(\kappa)$ at zero bias as free parameters. Figure~\ref{Fig:EigenfreqG0S1} shows the obtained values of $\omega_0(\kappa)$ and compares them with theory.
As expected from Fig.~\ref{Fig:FMandAFMS1} we find very good agreement. As the finite injector size influences the eigenfrequency spectrum, Fig.~\ref{Fig:EigenfreqG0S1} also shows the theoretical dependences of $\omega_0(\kappa)$ for ideal discontinuities. Clearly, only for large discontinuities, \ie $\kappa\gtrsim\pi$, differences become significant.
%
%

In sample 1 only the lowest eigenmode could be investigated due to limitations of our rf-generator. To experimentally investigate higher eigenmodes we measured sample 3, which is the longest JJ and also has the lowest plasma frequency. Figures~\ref{Fig:FMandAFMS3}(a) and (b) show the observed resonance peak positions of the measured FM and AFM configurations. 
Parasitic resonances are not plotted. Interestingly, in the FM configuration both eigenmodes $\omega_\pm$ can clearly be seen, yet only the lowest mode is present in the AFM state. Further measurements confirmed, that for all symmetric FM states we selected, the observed resonances clearly showed two modes coinciding with the theoretical dependence of $\omega_\pm$, yet only a single mode for the AFM states. Even more surprising, measurements on two other samples of proper length (not shown) revealed no trace of $\omega_1$, neither in AFM nor the FM state, indicating some general problem with the excitation of higher modes --- sample 3 being the exception. So far it is unclear to us, whether this behavior is owed to the particular way we apply microwaves to the JJ, \ie by capacitive coupling to the bias line, or to a more fundamental reason.
%

%
\begin{figure}[tb] 
  \begin{center}
    \includegraphics{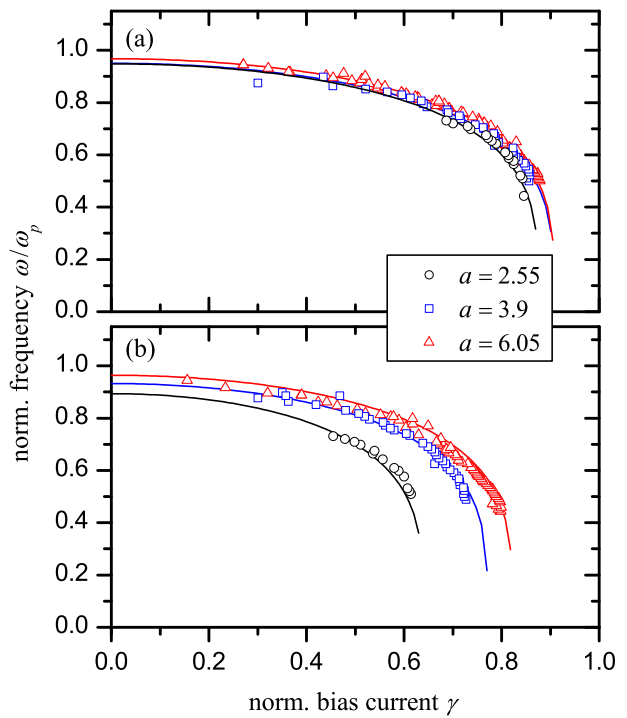}
  \end{center}
  \caption{%
    (Color online) Comparison of the resonance peak positions of the lowest mode for annular LJJs of different lengths with two fractional Josephson vortices: experiment (open symbols), numerical simulations (solid lines). (a) FM $(-0.6\pi,-0.6\pi)$-molecule. (b) AFM $(-0.6\pi,+0.6\pi)$-molecule.
  }
  \label{Fig:EFLangevgl}
\end{figure}
As has been shown in Fig.~\ref{Fig:EigenFreqOfa} coupling between the vortices in a molecule can be adjusted by changing the distance $a$ between them --- the smaller is the distance, the larger is the coupling (frequency splitting). Clearly, compared to the FM configuration, the deviation of the frequency of the lowest mode from the single vortex frequency~(\ref{Eq:SG}) is much more pronounced in the AFM case. The configuration related, different dependence of $\omega_0$ on $a$ can also be seen in Fig.~\ref{Fig:EFLangevgl}, where we directly compare $\omega_0(\gamma)$ of the FM and AFM ordered molecules $(\wp_1,\wp_2)=(-0.6\pi,-0.6\pi)$ and $(\wp_1,\wp_2)=(-0.6\pi,+0.6\pi)$ for three different vortex distances $a$. As expected, the eigenfrequency $\omega_0$ of the AFM state is much more sensitive to $a$ than of the FM state. 
Finally, we used Eq.~(\ref{Eq:omegagamma}) to obtain $\omega_{0,(1)}(\kappa)$ at $\gamma=0$ from measurements, which are shown in Fig.~\ref{Fig:EigenFreqOfa}. 

\section{Conclusions}
\label{Sec:Conclusions}
We experimentally investigated the interaction of fractional Josephson vortices in a symmetric two-vortex molecule by measuring the oscillatory eigenmodes of the molecule  spectroscopically. We observe a splitting of the single vortex eigenmode due to vortex-vortex coupling. The emerging molecule modes correspond to an in-phase and an out-of-phase oscillation of the individual vortices. The measured dependence of the eigenfrequencies of the molecule on applied bias current, topological charge $\wp$ of the vortices and vortex ordering are in good agreement with numerical calculations. We find that with decreasing vortex distance coupling and frequency splitting increases. Depending on the configuration of the molecule, \ie parallel or antiparallel vortex order, spectra show significantly different behavior, which can, in principle, be used to identify an unknown vortex state.   

\begin{acknowledgments}

This work was supported by the Deutsche Forschungsgemeinschaft via SFB/TRR-21 and the DFG Center of Functional Nanostructures (project B1.5). U.~K. gratefully acknowledges support by the Bisch\"{o}fliche Studienf\"{o}rderung Cusanuswerk, K.~B. and H.~S. by the Evangelisches Studienwerk~e.V.~Villigst.

\end{acknowledgments}


\end{document}